\documentstyle[prl,aps,multicol,epsf]{revtex}
\begin{document}
\title{Scaling Behavior of Cyclical Surface Growth}
\author{Yonathan Shapir$^{(1)}$, Subhadip Raychaudhuri$^{(1)}$, David 
G. Foster$^{(2,3)}$, and Jacob Jorne$^{(3)}$}
\address{$^{(1)}$Department of Physics and Astronomy, University of 
Rochester, Rochester, NY 14627\\
$^{(2)}$Eastman Kodak Company, Rochester, NY14650\\
$^{(3)}$Department of Chemical Engineering, University of 
Rochester, Rochester, NY 14627}
\date{\today}
\maketitle
\begin{abstract}
The scaling behavior of cyclical surface growth 
({\it e.g.} deposition/desorption), with the number of
cycles $n$, is investigated. The roughness of surfaces grown  
by two linear primary processes follows 
a scaling behavior with asymptotic exponents 
inherited from the dominant process while the effective amplitudes are
determined by both. Relevant non-linear effects in the primary processes 
may remain so or be rendered irrelevant. Numerical simulations for several 
pairs of generic primary processes confirm these conclusions. Experimental 
results for the surface roughness during cyclical electrodeposition/dissolution 
of silver show a power-law dependence on $n$, consistent with the scaling 
description.
\end{abstract}
 

Much interest has been devoted in recent years to scaling phenomena
in kinetic growth of self-affine surfaces. They are observed in a variety
of phenomena such as crystal growth, vapor deposition, molecular-beam
epitaxy, electrochemical processes, bacterial growth, burning fronts, etc.\cite
{family1} \cite{bara} \cite{villain} \cite{dkim}.
Similar rough surfaces may also be generated by reciprocal processes of
surface withdrawal caused by desorption, corrosion, evaporation, 
dissolution, and alike \cite{iwamoto}.
	In many natural and artificial systems of interest, however,
surfaces are not formed by a single process of growth or a sole process of
recession. Rather, they are the product of a combination of both. We
investigate here the fundamental scaling properties of surfaces formed by
cyclical growth processes. We focus on surfaces formed by two alternating
primary processes. Growth/recession cycles  (excluding the trivial 
cases in which the primary processes are time-reversed images of each other) 
are our main interest but cycles of two different growth (growth/growth) 
processes will be addressed as well. Cyclical behavior, prevailing 
in all natural phenomena, may be found in 
many of the systems cited above. Typical examples include weather and light
affected processes in organic (as the expansion/curtailment of a grass lawn
or a bacterial colony according to the availability of water or 
nutrient)
or non-organic (like alternating underwater erosion/sedimentation whether
or not the water is flowing) systems. They are also widespread in  
technological applications (such as rechargeable batteries for which shortcircuit 
by the metal accumulated on the electrodes is one of the failure 
mechanisms). Understanding the cyclical scaling properties 
may lead to accelerated testing, and performance improvement, of such 
systems.

	Our main challenge is to generalize the scaling approach in order
to make it applicable to cyclical growth processes. The
analytical, numerical, and experimental investigations, summarized below,
lead us to the conclusion that this is indeed possible, provided that
{\it the number of cycles $n$ substitutes for the time variable $t$ 
in the scaling relations.} We explore this generalized scaling behavior 
in several cyclical processes and reach some general conclusions on 
their behavior and their relations to the scaling relations of the primary
processes, of which they are composed.

	In the  primary processes, the surface width $W(L,t)$, where $L$ is the
lateral size of the system, is defined as
$ W(L,t) = <(h( \vec{r},t) - h(t))^{2}> ^{\frac{1}{2}}$. In this
definition $h(r,t)$ is the surface height and  $h(t)=<h(t)>=vt$ is the average
height, with $v$ being the average growth velocity.  $W(L,t)$  obeys the
following scaling form\cite{family1}\cite{family2}:

\begin{equation} \label{eq:1}
 W(L,t) \sim  L^{\alpha}g(L/\xi(t)),
\end{equation}
$\xi(t) \sim  t^{1/z}$ is the lateral correlation length. For large time
$t >> L^{z}$: $W \sim L^{\alpha}$,
while for $t<< L^{z}$:  $W \sim t^{\beta}$,  where $\beta = \alpha
/z$ is the growth exponent.
A related relation holds for the mean square height difference
$ <(h(\vec{r}+\vec{x},t)-h(\vec{x},t))^{2}> = 2(C(0,t)-C(r,t))$,
where $C(\vec{r},t)$  is the (equal-time) height-height correlation function.
 
The growth processes fall into different universality 
classes\cite{bara}. All
processes within one class share the same exponents and their  asymptotic
continuum stochastic equations differ at most by irrelevant terms (in the
renormalization group (RG) sense).
Using the symbolic  index $i=1,2,..$ to denote  different  processes, the
ones we consider here follow growth equations of the form:

\begin{equation} \label{eq:2}
\frac{\partial h(\vec{r},t)}{\partial t} = A_{i} \{h\} +
\eta_{i}(\vec{r},t) + v_{i},
\end{equation}
where $A_{i}$\{$h\}$ is a local functional depending on the spatial
derivatives of $h(\vec{r},t)$ and $<\eta_{i}(\vec{r},t)\eta_{i}(\vec{r}',t')>=2D_{i}
\delta(\vec{r}-\vec{r}')\delta(t-t')$.

	We recall the simplest generic growth processes of this type: (i)
Random deposition (RD) for which $A_{RD}=0$  and $\beta = 1/2$ ($\alpha$ and
$z$ are not defined);
(ii) The Edwards-Wilkinson (EW) model \cite{edwards} of preferred
growth at local minima has $A_{EW}= \nu {\mathbf \nabla}^{2}h$ and $\alpha=
\frac{3-d}{2}$, $\beta=\frac{3-d}{4}$, $z=2$;
(iii) The Kardar-Parisi-Zhang (KPZ) \cite{kardar} model which accounts for the growth being
locally normal to the surface has $A_{KPZ}= \nu {\mathbf\nabla}^{2}h +
 \frac{\lambda}{2}({\vec\nabla}h)^{2}$ and	$\alpha=1/2$,
$\beta=1/3$ and $z=3/2$ in 2d, while
$\alpha \simeq 0.39$, $\beta \simeq 0.24$ and $z \simeq 1.61$ in $3d$;
(iv) The Das Sarma-Tamborenea (DT)\cite{dasarma} (see also\cite{wolf}) for 
MBE deposition or
growth on kink sites has $A_{DT} = \nu_{4}{\mathbf \nabla}^{4}h$ with
$\alpha=\frac{5-d}{2}$, $\beta=\frac{5-d}{8}$, and $z=4$.

	We investigate theoretically and experimentally the  hypothesis that
in cyclical processes the scaling law in Eq. (1) should be replaced with:
\begin{equation} \label{eq:3}
W_{c}(L,n) \sim L^{\alpha}g_{c}(L/\xi_{c}(n))
\end{equation}
with $\xi_{c}(n) \sim n^{1/z}$.

	We begin with the study of cyclical processes composed of two
primary linear processes, namely those for which $A_{i}\{h\} =
a_{i}(\vec{\nabla})h(\vec{r},t)$,
where $a_{i}(\vec{\nabla})$ is a differential operator ({\it e.g.} the EW and
DT models) and for which time-reversal symmetry holds if the height 
is defined w.r. to the average height ({\it i.e.} for $h(\vec{r},t)-h(t)$).
 The first process in the cycle ($i=1$) is of duration $T_{1}= pT$ and second
process ($i=2$) lasts $T_{2}=(1-p)T$. The total duration of one cycle is
$T=T_{1}+T_{2}$. 
 
We also define $f(t)$ as the fractional part of
$t/T$. The growth equation thus becomes:
\begin{equation} \label{eq:4}
\frac{\partial h}{\partial t}=[a_{1}h + \eta _{1}]\Theta(p-f(t)) +
[a_{2}h + \eta_{2}]\Theta(f(t)-p),
\end{equation}
where $\Theta(x)$ is the unit step function.

	For such linear processes, the full scaling behavior may be
retrieved by looking at
$h(\vec{q},t)$, the Fourier transform (FT) of $h(\vec{r},t)$. The first observation we
make is  that in these time-reversible processes, only the averaged height
$h(t)=h(\vec{q}=0,t)$ is sensitive to the difference between growth
($v_{i} >0$) and recession ($v_{i}<0$). In terms of the average 
velocity $v_{c} = pv_{1}+(1-p)v_{2}$, 
it is given by:

$h(t)/T= nv_{c} +v_{1}f(t)\Theta(p-f(t)) +
[(v_{1}-v_{2})p+v_{2}(f(t)]\Theta(f(t)-p)$.

The roughness is insensitive to the sign of $v_{i}$ and hence will not discern
between a growth/growth and a growth/recession cyclical process (as long as
$a_{i}$ and $D_{i}$ are not altered).

In Fourier space, the growth equations for the modes with $\vec{q} \neq 0$ may be
integrated.
The $h(\vec{q},t)$ after $n$  ({\it res.} ($n+p$)) cycles are assigned as the initial
conditions for the $(n+1)^{th}$ application of the first ({\it res.} second) process.
The structure factor 
$S(q,t) = <h(q,t)h(-q,t)>$ (FT of $C(r,t)$, for simplicity we assume spatial isotropy in the 
basal plane) is then derived by
averaging over the noise. 
 We define $\bar{a_{i}} = a_{i} (q)T_{i}$  and  $\bar{a}_{c} = [a_{1}p +
a_{2}(1-p)]T$, in terms of which we find $S_{c}(q,n) \equiv S(q, nT)$ after
exactly $n$ cycles to be:

\begin{equation} \label{eq:5}
S_{c}(q,n)= \exp$\{$-2\bar{a}_{c}n$\}$ S(q,0)+
\left[\frac{D_{1}}{a_{1}}\exp(-2\bar{a}_{2})
(1-\exp(-2\bar{a_{1}}))
+\frac{D_{2}}{a_{2}}(1-\exp(-2\bar{a}_{2}))\right]
\left[{1-\exp(-2\bar{a}_{c}n)\over1-\exp(-2 \bar{a}_{c})}\right].
\end{equation}
A similar expression may be obtained for $S_{c}(q, n+p)$ and
straightforwardly extended to any time $t=(n+f)T$.

For small $q$, s.t. $\bar{a}_{c}(q)<<1$, $S_{c}(q,n)$ takes the form:

\begin{equation} \label{eq:6}
S_{c}(q,n) \sim
\frac{D_{c}}{a_{c}(q)}
\left[1-\exp(-2(a_{c}(q)Tn)\right],
\end{equation}
where we introduce the {\it effective} parameters of the cyclical growth
process:
$D_{c}=pD_{1}+(1-p)D_{2}$, 
and  $a_{c}(q) = pa_{1}(q)+(1-p)a_{2}(q) = \bar{a}_{c}/T.$
The same coclusion may be reached from coarse graining Eq.(4) and 
eliminating all modes with frequencies larger than $2\pi/T$\cite{subha} 
(this also yields the corresponding propagator $G_{c}(q,n)$ 
on time scales larger than $T$).
In terms of these effective parameters, $S_{c}(q,n)$ ($G_{c}(q,n)$),
with large $n=t/T$, 
of the cyclical process are equivalent
to $S(q,t)$ ($G(q,t)$) of a primary linear process. Hence,
the scaling behavior, presumed for the former in Eq.(3), indeed 
substitutes for that of Eq.(1) which holds for the latter\cite{bara}.

In the $n \rightarrow \infty$ limit, the asymptotic large-sacle 
roughness is determined by the small $q$ divergence of $S_{c}(q,t)$. Since
$a_{i}(q)\sim |q|^{z_{i}}$, it is the  process with the smaller $z_{i}$ which 
dominates the large $L$ cyclical roughness: $W_{c}(L,\infty)=AL^{\alpha}$, 
with $\alpha = min(\alpha_{1}, \alpha_{2}) = min(z_{1}, z_{2}) - (d-1)$   
(the {\it amplitude A} is proportional to  $D_{c}$ and is determined 
by {\it both} primary processes). The larger $\alpha_{i}$ appears as a correction to
scaling exponent (whether or not it is the leading one depends on how its
contribution compares with that of a potential subleading term in the 
dominating $a_{i}(q)$).
Note that a subleading term in $a_{c}(q)$ might affect the behavior
on a smaller scale, if its amplitude is large. In that case, the leading
behavior takes over only beyond a crossover length (at which both
contributions are comparable). As could be expected, 
the longer the non-dominant process lasts, the larger is the
crossover scale to the dominant behavior.

As for the dynamic exponent $z$ of the cyclical process, since $n$ appears 
always multiplied by $\bar{a}_{c}(q)$, the slower of
the two primary processes dictates the cyclical dynamics: 
$z = min(z_{1}, z_{2})$. 
The correlation length increases with $n$ as:
$\xi_{c} = n^{1/z}$ with an amplitude proportional to $[T_{i}]^{1/z}$,
with $T_{i}$ of the dominating process. For the initial cycles $(nT<<L^{z})$, 
the roughness grows as $n^{\beta}$, with $\beta = \alpha/z = (z - (d-1))/z$.

For processes described by stochastic non-linear equations, the
asymptotic behavior is explored by the RG approach \cite{bara}\cite{kardar}. 
An approximate RG procedure 
for cyclical processes may consist in first coarse graining the free
cyclical propagator (obtained from the two primary free propagators as 
explained above) until it becomes that of an effective linear, 
non-cyclical
process. Using this effective linear process as the ``free'' part, 
all the  non-linear terms of the primary processes, with bare couplings 
multiplied by $p$ (or $1-p$), are added as ``interactions''. The RG flows 
may then be derived following the ususal steps\cite{subha}.
   In this approximation the initial flow of the 
couplings is replaced by a simplified one. It is implicitly assumed that 
simplfying the initial flow will not alter the ultimate fixed 
point for each of the renormalized couplings. 
Although this seems very plausible,
it might not hold for all cyclical processes. 
  
Assuming it does, some conclusions may be reached. We begin with only 
one of the primary processes possessing a relevant non-linear term.
Its importance for the cyclical growth will depend on its relevance
with respect to the coarse grained
linear approximation of the cyclic propagator.
We may conclude that if the linear dominant term originates from
the same primary process as the non-linearity, the latter will remain
relevant. If, however, the non-linear process yields the non-dominant
part of the free propagator, the non-linearity may 
be rendered irrelevant, in which case the cyclical
behavior will be that of the other (linear) process.

If both primary processes have a non-linear term, a simple behavior is reached
if one of them is rendered irrelevant. Then the other one will
bequeath its scaling exponents to the cyclical descendant.
Theoretically, non-linear contributions from both processes 
may be relevant with various potential outcomes\cite{kpk}.

To examine these general conclusions, 
we performed numerical simulations on specific 
lattice atomistic models\cite{bara} in 2d (preliminary simulations in 3d show
a similar behavior). 
The system size in the simulations was changed between 128 to 4096 lattice 
spacings. A typical
cycle consisted of a deposition of 5-20 layers (average number of particle 
deposited per site) and desorption of
between 10$\%$ to 100$\%$ of the deposited amount. The maximum number of 
cycles $n$
varied between $500 - 10000$, averaged over 50-5000 independent runs,
depending on the pairs of primary processes 
and the system size. The 
growth exponent $\beta$ was extracted for different system size L. The value 
quoted is from the largest L (once $\beta$ became size-independent). 
From $W(L,\infty)$, the saturation width 
dependence on $L$, the roughness exponent 
$\alpha$ was derived. In most cases we checked independently 
the value of $\alpha$ from the scale dependence of the rms height-difference.

For linear primary processes,
we looked at the pairwise combinations of RD/EW, its reverse EW/RD,
and DT/EW, using the standard 
absorption/desorption algorithms for RD\cite{vold}, EW\cite{family3}, and 
DT\cite{dasarma}\cite{kim} (see also \cite{bara}). They all showed 
asymptotic cyclical 
scaling with EW exponents. This confirms the above conclusions for primary 
linear processes since EW is the dominating one when paired with RD or DT. 
We also run the respective adsorption/adsorption cycles  
using the same pairs. The roughness behavior was identical,
while their growth velocities were naturally different.

  To simulate non-linear processes we included two lattice realizations which 
  belong to the KPZ universality class :
 Ballistic Deposition (BD) \cite{vold}\cite{family2}\cite{meakin} 
 and the restricted SOS 
 algorithm of Kim and
 Kosterlitz (KK) \cite{kost}. We have generalized both of these 
 algorithms to desorption as well\cite{subha}. Both realizations yielded 
 equivalent behaviors 
 when combined with other processes (as they do in simple non-cyclical 
 adsorption or desorption) and we quote the ones obtained 
 using the BD (Ballistic Deposition 
 or Desorption) alogarithm.  If we combine the non-linear BD (or KK) alogarithm 
 with the EW (or DT) linear process
 to form a cyclical process, 
 the above considerations lead us to 
 expect KPZ exponents. Indeed, the KPZ free propagtor is equivalent to the 
 EW (and dominating over the DT) one.
 The exponents obtained are $\beta=0.311(5)$,
 $\alpha=0.51(1)$ for EW/BD , and $\beta=0.322(5)$, 
 $\alpha=0.50(1)$ for BD/EW.
These asymptotic exponents are 
 consistent with the KPZ $\beta=1/3$ (and, of course, with 
 $\alpha=1/2$,
 which is the common value of EW and KPZ). However,
 while for BD/EW these values were obtained for all sizes, for EW/BD
 the exponent $\beta$ increased slowly
 with the system size and the effective $\beta$ reached its
 asymptotic value only for the largest system size ($L=4096$).
 Growth/growth cycles yielded results very similar to the 
 growth/recession cycles. 
 
 To look at primary  processes with different values of $\alpha$, a 
DT $(\alpha_{1}=1.5)$ deposition with ballistic 
desorption $(\alpha_{2}=0.5)$ were performed. 
In Fig. 1 we show the logarithmic dependence of $W$ {\it (roughness)} on 
$\ln{(n)}$, for different system sizes $L$, for this DT/BD process. 
The inset depicts the logarithmic dependence of $Wm$
(the maximal value of the roughness)  {\it vs} $\ln{L}$.
From the graphs we obtain the asymptotic values of the
exponents for DT/KPZ:      
$\beta=0.311(15)$ and $\alpha=0.48(2)$, both consistent with the KPZ 
values.

Finally, simulations of BD/BD and KK/KK (note that they are not 
time-reversed images of each other because of the non-linearity) gave 
surfaces with KPZ scaling for 
$T_{1}\neq T_{2}$. For $T_{1} = T_{2}$, however, EW behavior was found. 
This follows
from the non-linear KPZ terms in the primary processes having the 
same magnitude but opposite signs. Hence they exactly cancel each other in 
the coarse grained growth equation yielding a EW one.

Experiments of cyclical growth were performed by metal 
electrodeposition/dissolution of silver.
Electrodeposition has been used in recent years to study the scaling  
behavior of surface growth 
(see \cite{tong}\cite{kaha}\cite{schmidt} and references therein).
To explore cyclical growth,
multiple electrodeposition/dissolution  cycles were carried out on 
initially vapor-deposited silver substrates,
ranging from 1 to 20 cycles. The plating solution contained 0.092 M 
AgBr (silver bromide),
$0.23 M (NH_{4})_{2}S_{2}O_{3}$ (ammonium thiosulfate), and 0.17 M 
$(NH_{4})_{2}SO_{3}$ 
(ammonium sulfite). Each cycle consisted of plating
for 5 minutes followed by 2.5 minutes of electrodissolution with a 
current density of 0.8 $mA/cm^{2}$. Up to 20
cycles have been performed and the roughness was examined by AFM after
$n$ full cycles and after the deposition part of the cycles (namely after 
$n+p$ cycles with $p=2/3$)\cite{dave}.
Fig. 2 shows a $log-log$ plot of the rms roughness vs. the cycle number. 
The data is consistent with a power-law scaling and the  
fit yields $\beta=0.48(5)$. For comparison $\beta=0.71$ for electrodeposition 
only under the same conditions\cite{dave}. 
Future experimental measurements (on the cyclical as well 
as on the primary processes) will allow more detailed scaling 
analysis and quantitative comparison of the scaling relations 
with theoretical predictions.

To summarize, the results of complementary studies of 
cyclical growth processes were presented, and show
them to be amenable to scaling 
analysis. The scaling description holds, provided the time 
variable is replaced by the number of cycles. This conclusion is 
supported by the initial experimental findings. We have derived the 
cyclical behavior of two alternating linear processes and outlined how the 
RG approach may be applied in presence of non-linear 
effects in the 
primary processes. In all systems we have studied analytically or 
numerically, the related exponents are unaffected
by the the cycle period T or the relative durations (p and 1-p) of the two processes.
One crucial question is how this behavior might be affected if 
the duration of the   
deposition (and/or desorption) phases are not uniform. 
We plan to address irregular intermittent growth/recession processes 
in the future\cite{subha}.
\smallskip

One of us (SR) is grateful to A. Kundagrami for his assistance in the 
simulations.
This research was supported by the NSF CMS-9872103 (JJ, SR, YS)
and by the Eastman Kodak Company (DGF).

\vfill\eject

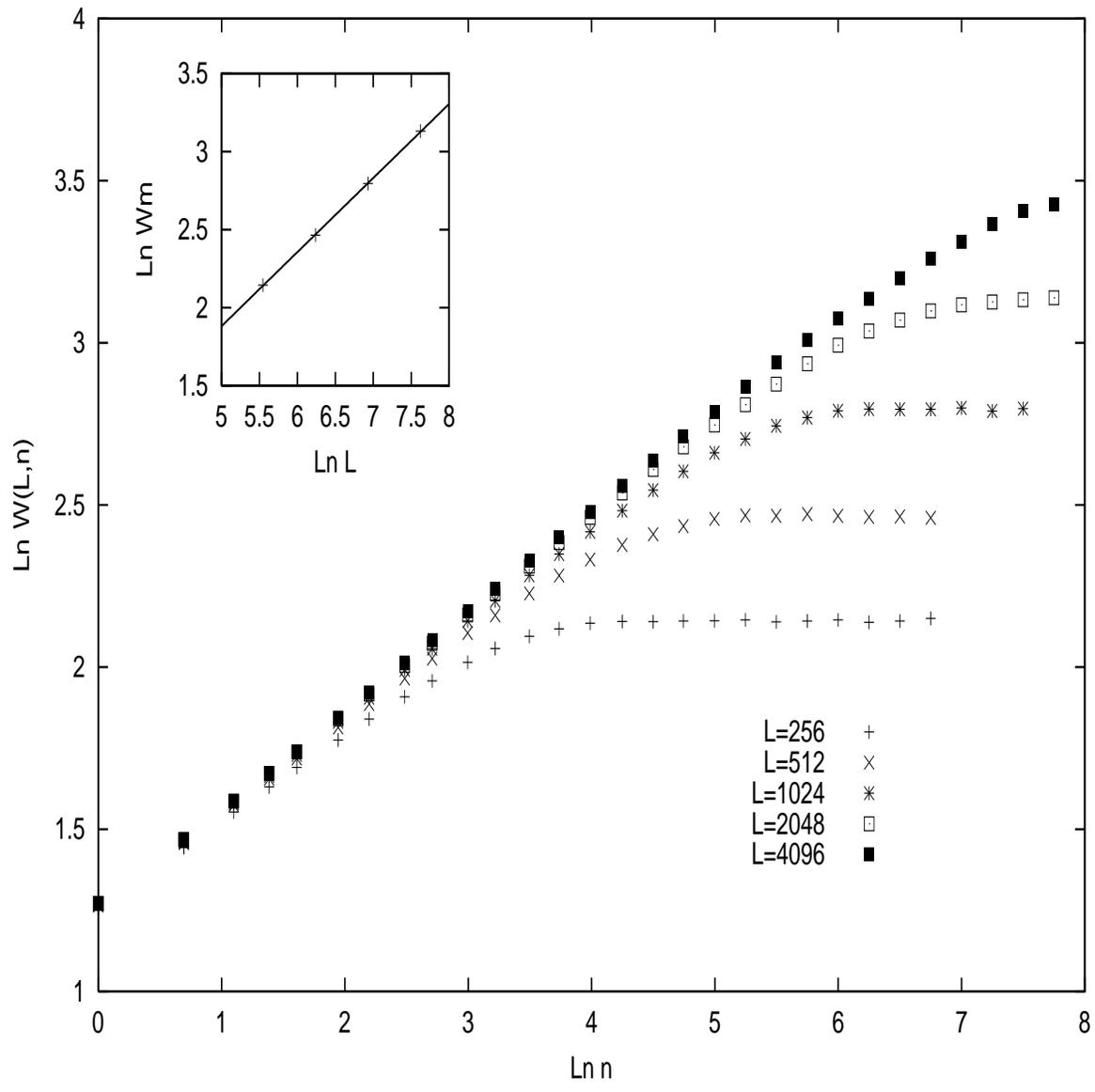
\begin{figure}
\caption{$\ln{W}$ (roughness) of the simulated cyclical DT/BD process {\it vs}  
$\ln{n}$ (number of cycles) for different system sizes $L$ 
(inset: $\ln{Wm}$ (maximal roughness) {\it vs} $\ln{L}$).}
\label{Fig. 1}
\end{figure}

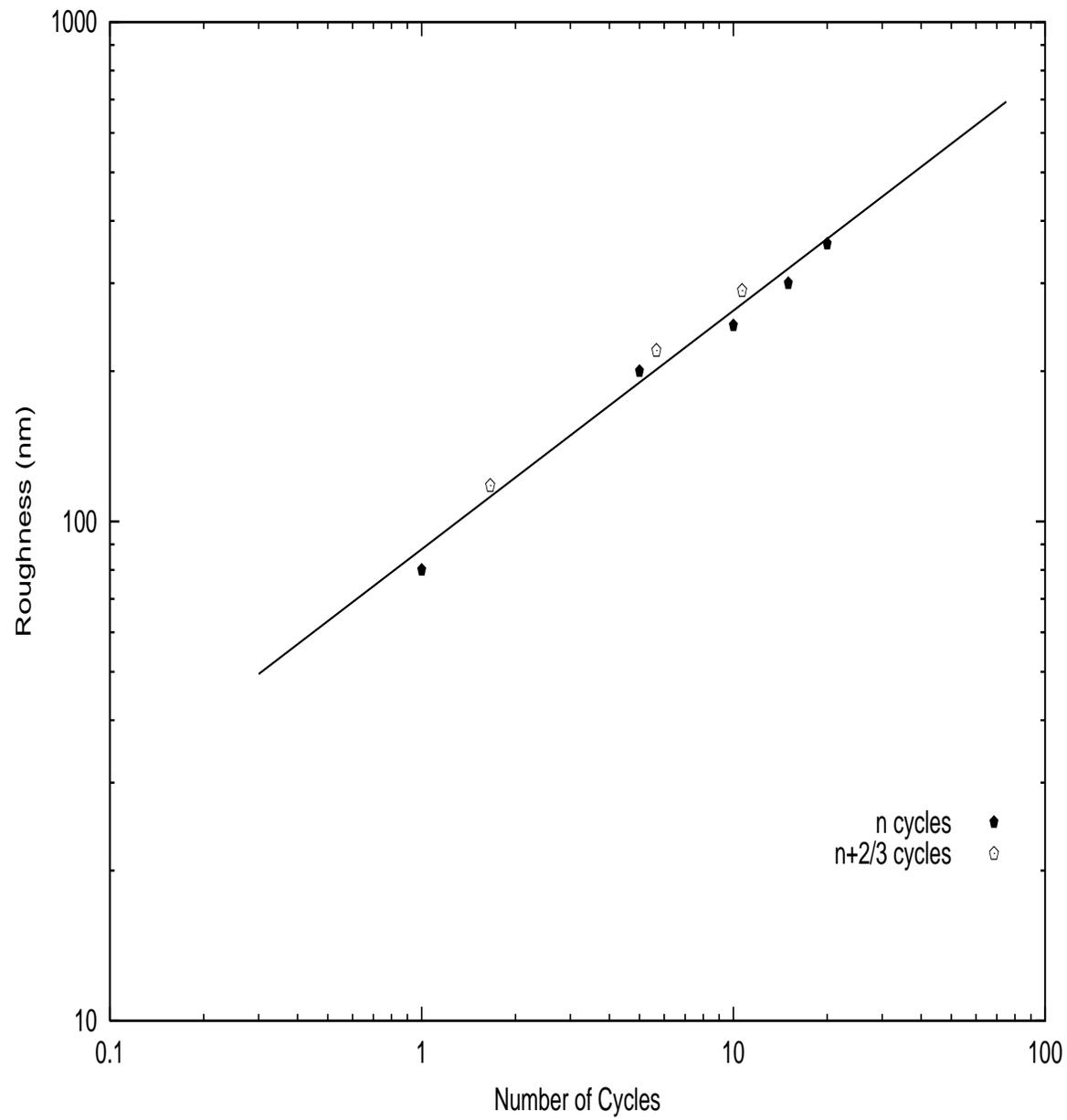
\begin{figure}
\caption{The roughness {\it vs} number of cycles $n$ in the 
electrochemical cyclical growth of Silver ({\it log-log} plot).}
\label{Fig. 2}
\end{figure}
 
\end{document}